# Electromagnetic cloaking by layered structure of homogeneous isotropic materials

**Ying Huang, Yijun Feng, Tian Jiang**

*Department of Electronic Science and Engineering, Nanjing University, Nanjing, 210093, CHINA*

*yjfeng@nju.edu.cn*

**Abstract:** Electromagnetic invisibility cloak requires material with *anisotropic* distribution of the constitutive parameters deduced from a geometrical transformation as first proposed by Pendry et al. [Science 312, 1780 (2006)]. In this paper, we proposed a useful method to realize the required radius-dependent, anisotropic material parameters and to construct an electromagnetic cloak through concentric layered structure of thin, alternating layers of *homogeneous isotropic* materials. With proper design of the permittivity or the thickness ratio of the alternating layers, we demonstrated the low-reflection and power-flow bending properties of the proposed cloaking structure through rigorous analysis of the scattered electromagnetic fields. The proposed cloaking structure does not require anisotropy or inhomogeneity of the material constitutive parameters usually realized by metamaterials with subwavelength structured inclusions, therefore may lead to a practical path to an experimental demonstration of electromagnetic cloaking, especially in the optical range.

## 1. Introduction

There is currently a growing interest in the theoretical and practical possibility of cloaking objects from the observation by electromagnetic fields [1-7]. The basic idea of the cloaking structure proposed by J. Pendry [1], is to use *anisotropic* transform medium whose permittivity and permeability are obtained from a homogeneous isotropic medium, by transformations of coordinates. The idea was successfully confirmed both by full-wave simulations [5] and by experimental demonstration at microwave frequencies [6]. However, the design use artificially structured metamaterial with inclusions of subwavelength metallic split-ring resonators (SRRs), and cannot be easily implemented for an optical cloak, which is certainly of particular interest. Recently, W. Cai *et. al.* have proposed an optical cloaking device for transverse magnetic (TM) polarization [7], but their design still require metamaterial with *anisotropic* distribution of the permittivity, which is realized by using subwavelength inclusions of metal wires in the radial direction embedded in a dielectric material.

Anisotropic materials with desired permittivity properties can be produced by a layered structure of thin, alternating dielectric layers (or metal and dielectric layers). Such planar systems have been proposed to demonstrate the subwavelength imaging [8], and to build photonic funnels for sub-diffraction light compression and propagation [9]. Recently, "optical hyperlens" made of cylindrical structure of anisotropic medium has been proposed, which has the capability of far-field imaging with resolution below the diffraction limit [10, 11]. Such anisotropic "hyperlens" has also been realized experimentally by concentric layered structures consisting of alternating layers of metal and dielectric [12, 13].

In this paper, we present the approach of realizing the electromagnetic cloaking by concentric layered structures. We show that by properly designing the realization of the anisotropic distribution of the permittivity required for the cloak through layered structure of *homogeneous isotropic* materials, the low-reflection and power-flow bending properties of the electromagnetic cloaking structure could be obtained.

## 2. Realizing anisotropic cylinder by layered structure of homogeneous isotropic materials

For the sake of simplicity, we restrict the problem to a two dimensional (2D) case. The proposed electromagnetic cloaking structure described here is based on the concentric layered structures consisting of alternating layers A and B of different homogeneous isotropic dielectric materials, as depicted in Fig. 1. When the layers are sufficiently thin compared with the wavelength, we can treat the whole layered structure as a single anisotropic medium with the dielectric permittivity as [8-10]

$$\varepsilon_\theta = \frac{\varepsilon_A + \eta \varepsilon_B}{1+\eta},$$

$$\frac{1}{\varepsilon_r} = \frac{1}{1+\eta}\left(\frac{1}{\varepsilon_A} + \frac{\eta}{\varepsilon_B}\right),$$
(1)

where, $\varepsilon_A$, $\varepsilon_B$ are the permittivities of the layer A and layer B, respectively, $\varepsilon_\theta$, $\varepsilon_r$ are the angular and radial components of the effective anisotropic permittivity tensor, respectively, and $\eta$ is the thicknesses ratio of the two layers:

$$\eta = \frac{d_B}{d_A}.$$
(2)

Treating the layered structure as an effective anisotropic medium is based on the effective medium approximation, which requires that the thickness of each layer is much less than the wavelength and the number of the layers is large enough. When considering the finite thickness of a practical layered structure, it has been demonstrated in a planar case that the effective-medium approximation becomes more appropriate as the layers are made thinner [8]. Here we give the electromagnetic analysis on the cylindrical case in the following.

Consider the electromagnetic wave scattering for an infinite conducting cylinder shelled either with a concentric layered structure (Fig. 1(a)) or with an equivalent anisotropic medium (Fig. 1(b)). A plane wave with TM polarization is assumed to impinge along the $x$ direction upon the shelled cylinder. The axial components of the incident and scattered magnetic field vector **H** outside the cylinder may be expressed as [14]

$$H_z^i = H_0 \sum_{n=-\infty}^{n=\infty} j^{-n} J_n(k_0 r) e^{jn\theta},$$

$$H_z^s = H_0 \sum_{n=-\infty}^{n=\infty} C_n j^{-n} H_n^{(2)}(k_0 r) e^{jn\theta},$$
(3)

where, $J_n$, $H_n^{(2)}$ are, respectively, the Bessel function and the Hankel function of the second kind, $k_0$ is the wave number of the outside medium. The magnetic field inside the $m^{th}$ dielectric layer (designated by the subscript $m = 1$ to $2N$) of the layered structure is expressed by

$$H_{zm}^t = H_0 \sum_{n=-\infty}^{n=\infty} j^{-n}\left(A_{mn} J_n(k_m r) + B_{mn} H_n^{(2)}(k_m r)\right) e^{jn\theta},$$
(4)

where, $k_m$ is the wave number in the $m^{th}$ dielectric layer. In Eq. (3) and (4), a time dependence of the form $e^{j\omega t}$ is assumed for the electromagnetic field quantities but is suppressed throughout. $A_{mn}$, $B_{mn}$, and $C_n$ are arbitrary constants that can be determined by enforcing the boundary continuity condition at $r = a$, $r = b$, and at the interfaces between each dielectric layer. The electric fields in each medium are determined through

$$E_r = -\frac{1}{r}\frac{j\omega\mu}{k^2}\frac{\partial H_z}{\partial \theta}, \quad E_\theta = \frac{j\omega\mu}{k^2}\frac{\partial H_z}{\partial r}. \tag{5}$$

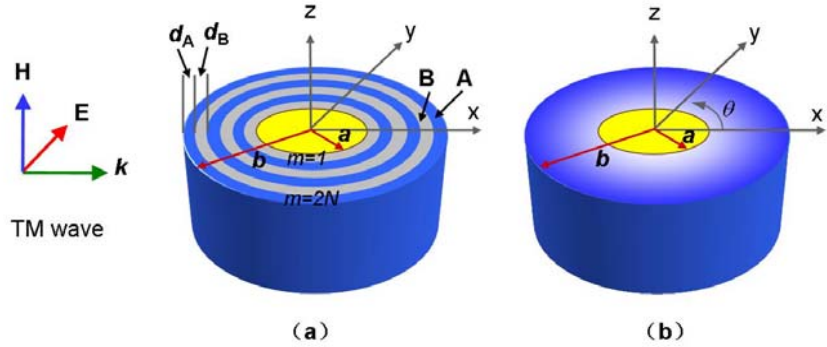

Fig. 1. TM wave incident on an infinite conducting cylinder (in yellow) shelled with (a) concentric layers structure with alternating layers of dielectric A and B, (b) equivalent anisotropic cylindrical medium with radius-dependent, anisotropic material parameters. Both of the shells have inner and outer radius of $a$ and $b$, respectively.

For a TM incident wave from a magnetic line source (with magnetic current of $I_m$, and located at $\mathbf{r} = \mathbf{r}_0$), the incident magnetic field is described as

$$H_z^i = -\frac{\omega\varepsilon_0 I_m}{4} H_0^{(2)}\left(k_0|\mathbf{r}-\mathbf{r}_0|\right). \tag{6}$$

The scattering electromagnetic fields can be analyzed similar to that of the plane wave case.

When the conducting cylinder is shelled with the equivalent anisotropic medium (Fig. 1(b)), the electromagnetic fields scattering could be calculated similarly by application of Sommerfeld's bundle of rays field representation via a polarization dependent coordinate transformation [15]. The electromagnetic wave scattered by the cylinders could be verified by calculating the far-field scattering pattern which is proportional to

$$\xi(\theta) = \left|\sum_{n=-\infty}^{n=\infty} C_n e^{jn\theta}\right|^2. \tag{7}$$

First, we calculate the scattering fields by the conducting cylinder shelled with $2N$ layers of alternating dielectric A and B with same thickness ($\eta = 1$) as depicted in Fig. 1(a). We choose $\varepsilon_A = 7.46$, $\varepsilon_B = 0.54$, $a = \lambda$, $b = 2\lambda$, where $\lambda$ is the wavelength of the incident electromagnetic wave. The far-field scattering pattern is calculated and plotted in Fig. 2 for the shells composed of different numbers of alternating dielectric A and B ($N = 5$, and 20). The scattering pattern is also plotted in Fig. 2 for the conducting cylinder without any shell for comparison. Next, we calculate the scattering fields by the same conducting cylinder shelled with the equivalent anisotropic medium of same thickness, which has a 2D anisotropic permittivity of $\varepsilon_r = 1.0$, and $\varepsilon_\theta = 4.0$, determined by Eq. (1). The far-field scattering pattern is plotted and compared in Fig. 2. It shows clearly that as the layers are made thinner by increasing the number of layers from 10 to 40, the layered structure shell has nearly the same scattering property as the equivalent anisotropic medium shell. This means that we are able to use a concentric layered structure to realize a cylindrical shell of a 2D rotationally invariant anisotropic medium. The thinner the layers, the better this layered structure approaches the scattering performance of the anisotropic medium.

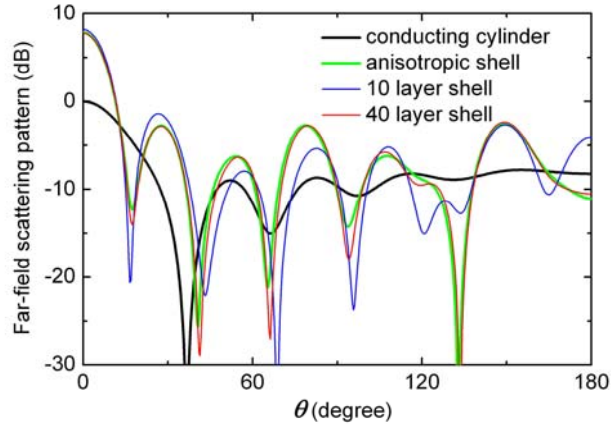

Fig. 2. Comparison of the far-field scattering patterns for a bare conducting cylinder, the conducting cylinder shelled with a concentric layered structure of different number of layers, and that shelled with an anisotropic cylindrical medium. All values have been normalized to the scattering pattern of the bare conducting cylinder at $\theta = 0$.

**3. Electromagnetic cloaking properties of layered structures**

As proposed in Ref. [5, 6], cloaking a central cylindrical region of radius *a* by a concentric cylindrical shell of radius *b* requires a cloaking shell with the following radius-dependent, anisotropic relative permittivity and permeability:

$$\varepsilon_r = \mu_r = \frac{r-a}{r}, \quad \varepsilon_\theta = \mu_\theta = \frac{r}{r-a},$$
$$\varepsilon_z = \mu_z = \left(\frac{b}{b-a}\right)^2 \frac{r-a}{r}.$$
(8)

Here we follow the TM wave illumination as considered in [7]. In this case only $\mu_z$, $\varepsilon_\theta$, and $\varepsilon_r$ are of interest and must satisfy the requirement of Eq. (8). Similar to the consideration in [6, 7], to make only one component spatially inhomogeneous and also to eliminate any infinite values in Eq. (8), we can choose a reduced set of medium parameters as

$$\mu_z = 1, \quad \varepsilon_\theta = \left(\frac{b}{b-a}\right)^2,$$
$$\varepsilon_r = \left(\frac{b}{b-a}\right)^2 \left(\frac{r-a}{r}\right)^2.$$
(9)

This allows us to completely remove the need for magnetic response of the material, which is especially important for making cloak at optical frequency. The shell with the reduced set of parameters provides the same wave trajectory inside the cloaking medium, but it will induce some unfavorable reflection at the outer boundary due to the impedance mismatch.

Now we consider the realization of the above radius-dependent, *anisotropic* shell through layered structures of *homogeneous isotropic* materials. As verified in the previous section, an anisotropic cylindrical shell (with $\varepsilon_\theta$, $\varepsilon_r$ as the angular and radial permittivity components) could be mimicked by concentric layered structure consisting of alternating layers of two different dielectrics (with $\varepsilon_A$, $\varepsilon_B$ as the permittivity), and the relation of the material parameters between them could be determined by Eq. (1), if only the thickness of each layer is much less compared with the wavelength. For an anisotropic shell with radius-dependent $\varepsilon_r$, we could still use a layered structure to mimic it by enforcing the permittivity of the alternating layers to vary with the radius. If the permittivity of the neighboring layers is changing gradually, we will show in the following examples that it is a good approximation to follow Eq. (1) to design the permittivity of the alternating dielectrics in the layered structure.

We consider the following design to explore the realizability of the cloaking through layered structures. For an example, we assume a cylindrical cloaking shell with inner radius *a* = λ and outer radius *b* = 2λ. Similar to other practical realization of the invisibility cloaks [5-7], we firstly consider a stepwise homogeneous *N*-layer approximation of the ideal continuous parameters required by Eq. (9), and the continuous radius-dependent, anisotropic medium (as shown in Fig. 1(b)) could be represented approximately by *N* discrete layers of homogeneous anisotropic medium. Then we mimic each homogeneous anisotropic layer by an alternating layers of isotropic dielectric A and B (as shown in Fig. 1(a)), and the permittivity of the two layers are designed by Eq. (1) from the corresponding anisotropic layer. Fig. 3 depicts the material parameters of the proposed cloak of alternating layers of isotropic dielectric A and B, each with *N* = 20 layers and of equal thickness ($\eta = 1$). The relative permittivity of dielectric A increases gradually with the radius of each cylindrical layer from 0.005 to 0.536, while the relative permittivity of dielectric B decreases gradually from 8.00 to 7.46.

To demonstrate the performance of the designed electromagnetic cloak, we assume a TM plane wave incidence along the *x* direction, and calculate the electromagnetic wave scattered from a conducting cylinder (with radius $a = \lambda$) shelled with the proposed cloak using the semi-analytical method described in the previous section. The calculated results of magnetic-field distribution around the cloaked cylinder together with the power flow lines are illustrated in Fig. 4. As shown in Fig. 4(a), with the cloaking structure the wave fronts are deflected and guided around the cloaked region, and return to the original propagation direction with very small perturbation. When remove the cloaking structure (Fig. 4(b)), the waves are severely scattered by the object resulting in a remarkable backward reflection and an evident shadow cast behind the conducting cylinder.

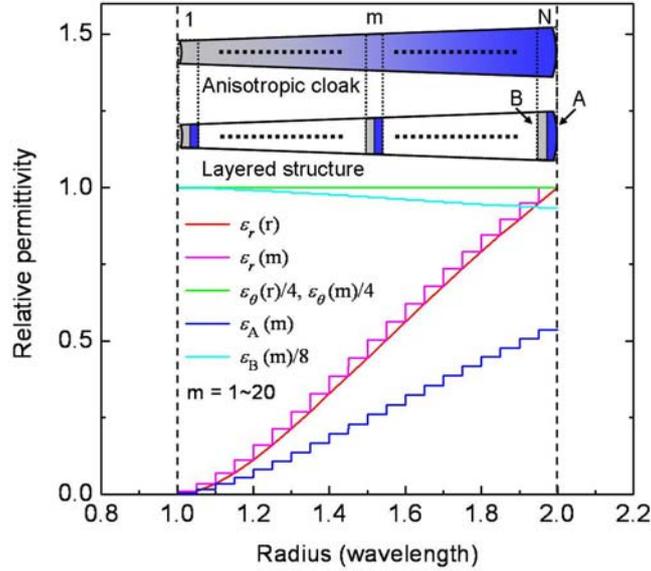

Fig. 3. The relative permittivity components required for an ideal reduced set of parameter ($\varepsilon_r(r)$, $\varepsilon_\theta(r)$), and that for the corresponding layered structure with alternating dielectric A and B ($\varepsilon_A(m)$, $\varepsilon_B(m)$). The inset describes the anisotropic shell divided into stepwise homogeneous N-layer (with permittivity as $\varepsilon_r(m)$, $\varepsilon_\theta(m)$), and the mimic of each layer by alternating layers of dielectric A and B (totally 2$N$ layers).

Due to the reduced set of parameters given by Eq. (9), there is an unavoidable but low level of reflection observed in Fig. 4(a), which is caused by the impedance mismatch at the outer boundary. The power reflection is about 11% estimated from the ratio of the inner and outer radii, $R_{ab} = a/b = 0.5$ [7]. This agrees with the backward reflection deduced from the standing wave observed in Fig. 4(a). Decreasing $R_{ab}$ or using a thicker cloak can further reduce the backward reflection. For example, if we choose $a = \lambda$ and $b = 3\lambda$, we have obtained less scattering from the cloak shell with a reduced power reflection of about 4%.

The above calculations have clearly demonstrated the capability of reducing the scattering from the conducting object inside the cloak and the power-flow bending properties of the proposed electromagnetic cloak. Such a concentric layered structure could get rid of the requirement of radius-dependent, anisotropic distributed relative permittivity usually realized by metamaterials with subwavelength structured inclusions. The most concerned issue of the proposed layered structure is the requirement of the alternating dielectrics with gradually increased or decreased permittivity, which is not easy to realize and to control from a practical point of view. An alternative approach is to construct a concentric layered structure that the permittivity of the alternating dielectrics is fixed, while the thickness ratio of the two layers varies to approximately satisfy the Eq. (9), since it will be easier to control the thickness of the dielectric layers.

For an example, a layered structure with $N = 10$ has been designed with $a = \lambda$ and $b = 2\lambda$. The relative permittivity of the alternating layers is assumed to be $\varepsilon_A = 0.01$, $\varepsilon_B = 4$, and the thickness ratio $\eta$ of the two layers varies from 2.3 to 130 according to Eq. (9). To verify the performance of the cloak, we assume a TM incident wave from a line source and calculate the electromagnetic wave scattered from a conducting cylinder (with radius $a = \lambda$) shelled with the proposed cloak. The magnetic-field distribution is plotted in Fig. 5 (a). Although there is imperfectness in the scattered fields caused by the reduced set of parameters and small approximation made in the design of the permittivity and the thickness ratio, Fig. 5(a) clearly demonstrates the low-scattering, shadow-reducing and power-flow bending properties of the electromagnetic cloaking structure. The electromagnetic wave scattering is also compared in Fig. 5 (b) for the non-shelled conducting cylinder irradiated by same the line source.

Materials with permittivity close to zero required for dielectric A in this design could be available at infrared and visible range. For examples, the permittivity of the noble metals and polar dielectrics follows Drude or Drude-Lorenz dispersion models [16, 17], and the real part of their permittivity effectively goes to zero at their plasma frequency, which, usually lies in the terahertz regime for polar dielectrics and some semiconductors [18] and in the visible and ultraviolet for noble metals [16-19]. Composite materials realized by properly embedding metallic nanoparticles and nanowires in a dielectric medium could also have effective permittivity near zero in a wide range of frequencies up to the visible [20, 21].

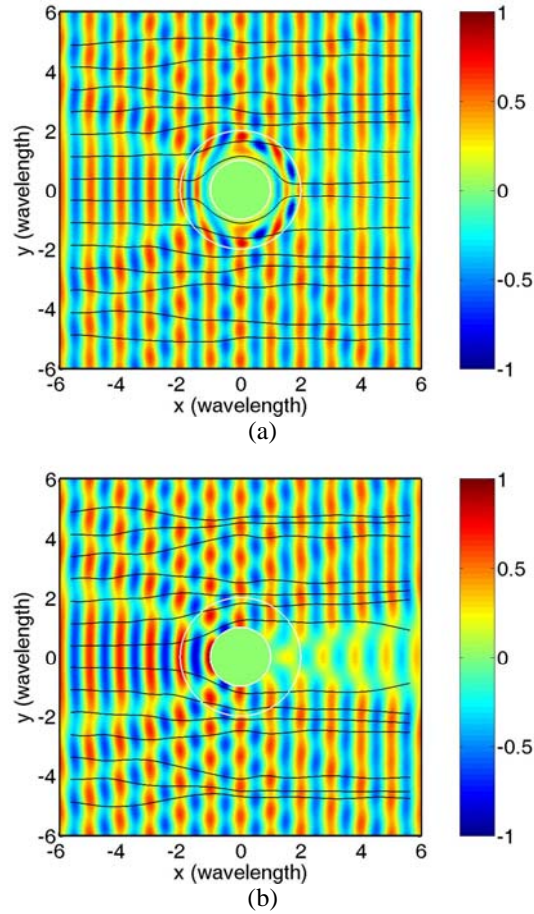

Fig. 4. The calculated magnetic-field distribution around the conducting cylinder (a) with a cloak of concentric layered structure, and (b) without cloak. Power-flow lines (in black) in (a) show the smooth deviation of electromagnetic power around the cloaked object. The white circles outline the cloak.

## 4. Conclusions

We have proposed an electromagnetic wave cloak by realizing the radius-dependent, anisotropic material through layered structures of homogeneous isotropic materials. The performance of the cylindrical cloak has been demonstrated through rigorous calculation of the electromagnetic wave scattering, which reveals the low-scattering and power-flow bending nature of the properly designed cloaking structure. Our proposal has no requirement of any anisotropy or inhomogeneity of the material constitutive parameters, therefore may lead to a simpler path to an experimental demonstration of electromagnetic cloaking, especially in the optical range.

**Acknowledgments**

This work is supported by the National Basic Research Program of China (2004CB719800), the

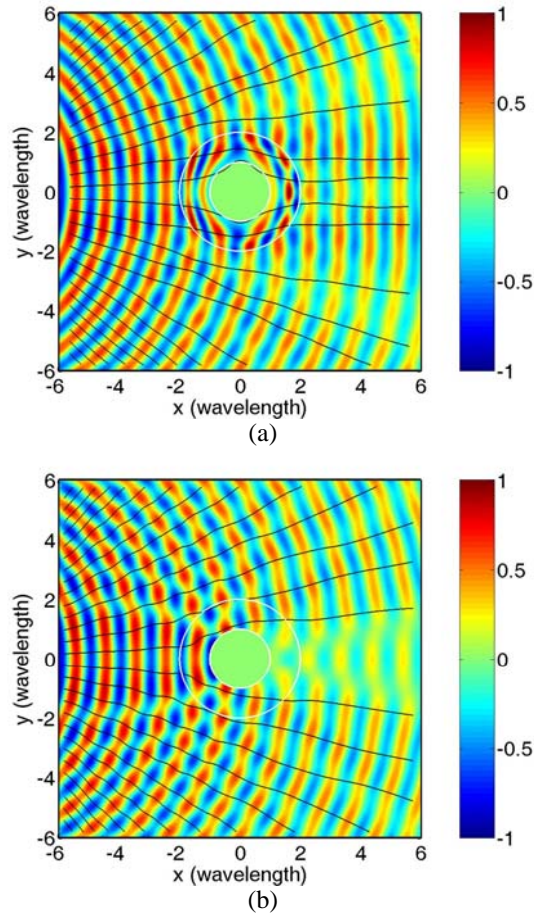

Fig. 5. The magnetic-field distribution around the conducting cylinder, (a) with a cloak of layered structure, and (b) without cloak, for a TM incident wave from a line source. The white circles outline the cloak.